\documentclass{article}
\usepackage[utf8]{inputenc}
\usepackage{amsmath}
\usepackage{amsfonts}
\usepackage{authblk}

\title{Chiral Dirac Equation and Its Spacetime and CPT Symmetries}
\author[1]{Timothy B. Watson}
\author[1]{Zdzislaw E. Musielak}
\affil[1]{University of Texas at Arlington}
\date{}   

\begin{document}
\maketitle

\begin{abstract}
The Dirac equation with chiral symmetry is derived using the irreducible representations of the Poincar\'{e} group, the Lagrangian formalism, and a novel method of projection operators that takes as its starting point the minimal assumption of four linearly independent physical states. We thereby demonstrate the fundamental nature of this form of the Dirac equation. The resulting equation is then examined within the context of spacetime and CPT symmetries with a discussion of the implications for the general formulation of physical theories.
\end{abstract}

\section{Introduction}

Since its introduction in 1928 numerous modifications of the Dirac operator [1] have been proposed resulting in a number of
so-called generalized Dirac equations. Such generalizations have been invoked for the purposes of 
unifying leptons and quarks [2-4], accounting for the three families of elementary particles [5-9], including ad hoc a pseudoscalar mass [10], and extending the Dirac equation to distances comparable to the Planck length [11]. One most recent modification of the Dirac equation has included the effects of chiral symmetry [12]. In this generalization, the most general first order partial differential equation for bispinors constrained via Poincar\'{e} invariance is derived, resulting in the chiral Dirac equation (CDE) which differs from the original by an additional degree of freedom in the form of the chiral angle. Specific solutions to this equation necessitate a specific choice of chiral basis. These results have been used to explain the smallness of neutrino masses and some properties of dark matter [12].

With the benefit of hindsight and the accumulation of experimental evidence, it may be said that one of the great insights of Driac was the observation that elementary particles exhibit a quantization of both angular momentum (spin-up/spin-down) and sign of the energy (matter/antimatter). Modern particle physics has reinforced this picture for free particles of a given flavor independent of internal (gauge) symmetries. In this work, we demonstrate how the assumption of two quantized characteristics is equivalent to the chiral form of the Dirac equation (CDE), Dirac's original equation being a special case.

In an attempt to demonstrate these result holistically, the present paper is organized as follows: Section 2 presents three different methods of deriving the CDE operator beginning with the standard group theory and Lagrange formalism derivations before presenting a novel derivation from the assumption of the existence of independent physical states and their corresponding orthogonal idempotents (projection operators) paramaterized by physically meaningful quantities; Section 3 then sets out a detailed investigation into the spacetime (continuous) and CPT (discrete) symmetries underlying the derived CDE, their role and constraints in the derivation procedures, and the physical implications resulting from this equation; Section 4 is devoted to conclusions. 
\section{Methods to Derive Chiral Dirac Equation}
\subsection{Group Theoretical Derivation}
To derive the Dirac equation with chiral symmetry, we begin by following Wigner [13,14] and identifying particles with the induced irreducible representations (irreps) of the Poincar\'{e} group $\mathcal {P}\ =\ SO (3,1) \otimes_s T (3+1)$, with $SO(3,1)$ being a non-invariant Lorentz group of rotations and boosts and $T(3+1)$ an invariant subgroup of spacetime translations. The condition that the Dirac spinor wavefunction transforms as one of the irreps of $T(3+1) \subset \mathcal {P}$ can be written as the following matrix eigenvalue equation [12]

\begin{equation}
X^{\mu} \partial_{\mu} \psi = - Y \psi\ ,
\label{eq1}
\end{equation}
where $\mu$ = $0$, $1$, $2$ and $3$ and

\begin{equation}
\psi =  \begin{bmatrix}
    \chi_{L} \\
    \chi_{R}\\
\end{bmatrix}
\label{eq2}
\end{equation}
is the Dirac spinor wavefunction with $\chi_{L}$ and $\chi_{R}$ being two-component bispinors. Here $X^{\mu}$ and $Y$ are five $4 \times 4$ which nontrivially mix the spinor components. To determine the form of these matrices, we constrain them to satisfy covariant conditions under the Lorentz transformations of SO(3,1), and find the resulting Poincar\'{e} invariant chiral Dirac equation (CDE) [12] to be given by 

\begin{equation}
(i \gamma^{\mu} \partial_\mu - m e^{i \alpha \gamma^5})\psi = 0\ ,
\label{eq3}
\end{equation}
where $\alpha$ is the chiral angle. This angle appears as a necessary degree of freedom as it is unconstrained by the covariant transformation constraints. It may be noted also that this degree of freedom is a consequence of taking our bispinors to be eigenstates of CPT symmetry and not eigenstates of parity as is conventionally required in deriving the Dirac equation \footnote{We thank M. Efroimsky for bringing this point to our attention} [15].

Because this method of deriving the CDE is based explicitly on the symmetries of Minkowski spacetime, we may be certain the resulting equation is Poincar\'e invariant. Furthermore, The CDE correctly accounts for all four sets of states (spins up and down, matter and antimatter), and its solutions have important physical implications [12].  Being Poincar\'e invariant and local, the CDE satisfies the two basic characteristics required for the equation to be fundamental.  
Two other characteristics are gauge invariance and the existence of Lagrangian. Since we consider only free elementary particles, gauge invariance is not discussed here. We now demonstrate that the CDE can be formally derived from Lagrangian formalism, which is sufficient to call the CDE the fundamental equation of physics.

\subsection{Derivation from Lagrangian formalism}

The Lagrangian formalism is a powerful and independent way to derive a dynamical equation. The Lagrangian for the Dirac equation ($\alpha = 0 $ in Eq. \ref{eq3}) is very well-known and presented in textbooks (e.g., [15,16]) without derivation. In fact, the Lagrangian was not a part of the Dirac's original  paper where the equation first appeared [1].  An interesting attempt to obtain the Dirac Lagrangian is presented and discussed in [17].  Let us briefly review the main points of this attempt and then use them to obtain the Lagrangian for Eq. (\ref{eq3}).

In case $\alpha = 0$, Eq. (\ref{eq3}) reduces to the Dirac equation 
\begin{equation}
(i \gamma^{\mu} \partial_\mu - m)\psi = 0\ ,
\label{eq4}
\end{equation}
which describes a free, massive, non-chiral and spin 1/2 relativistic elementary 
particle [1, 15-17].

To obtain the Lagrangian density for this equation, we follow [17] and require that the Lagrangian is a hermitian, single-valued proper scalar or pseudo-scalar in $\psi$ and $\partial_{\mu} \psi$.  Since $\psi$ has the double-valued properties under 
rotations, the terms in the Lagrangians must have even numbers of $\psi$. The simplest proper scalar is $\bar {\psi} \psi$, where $\bar {\psi}$ is the Dirac adjoint. Now, the construction of a scalar kinetic term has to be done
with caution as $\partial_{\mu} \psi$ requires saturation of the index $\mu$, which cannot be done by another derivative since the result would be a second-order equation. Therefore, the Dirac matrices $\gamma^{\mu}$ are used to saturate the index $\mu$, and write the kinetic term as $i \bar {\psi} \gamma^{\mu} \partial_{\mu} \psi$. Since the physical units of the kinetic term are different than $\bar {\psi} \psi$, the latter must be multiplied by an inverse length dimension, which in natural units is mass. Then, the Dirac Lagrangian can be written in the following form

\begin{equation}
{\mathcal {L}_D} = {1 \over 2} \bar {\psi} ( i \gamma^{\mu} \partial_{\mu}
- m ) \psi - {1 \over 2} \bar {\psi} ( i \gamma^{\mu} {\overleftarrow 
{\partial_{\mu}}} + m ) \psi= 0\ .
\label{eq5}
\end{equation}
This is a fully symmetric form of the Lagrangian, which shows that when
evaluated along a stationary path the Dirac Lagrangian vanishes [17].
Both the Dirac equation and its Lagrangian are Poincar\'e invariant.

Using the above procedure, the Lagrangian for the CDE (see Eq. \ref{eq3}) 
can also be obtained and written as 

\begin{equation}
{\mathcal {L}_{CD}} = {1 \over 2} \bar {\psi} ( i \gamma^{\mu} \partial_{\mu}
- m e^{-i \alpha \gamma^5}) \psi - {1 \over 2} \bar {\psi} ( i \gamma^{\mu} 
{\overleftarrow {\partial_{\mu}}} + m e^{-i \alpha \gamma^5} ) \psi= 0\ .
\label{eq6}
\end{equation}
Similarly to ${\mathcal {L}_D}$, the Lagrangian is also fully symmetric,
hermitian and single-valued proper scalar or pseudo-scalar, and it vanishes when evaluated along a stationary path.  Moreover, the CDE and its Lagrangian are Galilean invariant.  By substituting ${\mathcal {L}_{CD}}$ into the Euler-Lagrange equation for 
variations with respect to $\bar {\psi}$, the CDE given by Eq. (\ref{eq3}) is obtained. This method of deriving the CDE is independent from the group theory derivation and serves to demonstrate that the equation satisfies a least-action principle requisite of any fundamental theory. Our derivation based on projection operators is now presented.

\subsection{Derivation from Orthogonal Idempotents}

We may define a set of projection operators operating on an $N$-dimensional complex vector space with any set of $N$-by-$N$ orthogonal idempotent matrices satisfying

\begin{equation}
\hat{P}_{i} \hat{P}_{j} = \hat{P}_{j} \hat{P}_{i}= \delta_{ij} \hat{P}_{i} \qquad 
\sum^{N}_{i=1} \hat{P}_{i} = 1\ .
\label{eq7}
\end{equation}

The total number of projection operators of a given vector space is maximally equal to the dimensions of the space considered. Therefore, for $N=2$, we may expand the 
most general operators acting on a spinor in terms of the Pauli matrices and the two-by-two identity matrix.  Let

\begin{equation}
\hat{P}_{1} = a_0 I_2 + \vec{\sigma} \cdot \vec{a}  \qquad
\hat{P}_{2} = b_0 I_2 + \vec{\sigma} \cdot \vec{b}\ .
\label{eq8}
\end{equation}

Then enforcing the orthogonal idempotent conditions, we obtain 
the following constraints

\begin{equation}
 a_0 =  b_0 = \frac{1}{2} \qquad
 \vec{a} \cdot \vec{a} = \vec{b} \cdot \vec{b} = \frac{1}{4} \qquad
 \vec{a} = -\vec{b}\ .
\label{eq9}
\end{equation}

Solving this we find two projection operators for our symmetry group whose degrees 
of freedom may be parameterized in terms of the unit vector $\hat{a}$.   We write 
these projection operators succinctly as

\begin{equation}
\hat{P}_{\pm}(\hat{a}) = \frac{1}{2}(I_2 \pm \vec{\sigma} \cdot \hat{a})\ .
\label{eq10}
\end{equation}

For a fixed $\hat{a}$, these operators allow us to define two types of objects in our two-dimensional vector space. For any such element $\chi \in \mathbb{C}^2$ we may define $\chi_\pm (\vec{a})\equiv \hat{P}_{\pm}(\hat{a}) \chi(\vec{a})$. It necessarily follows that

\begin{equation}
(\vec{\sigma}\cdot \hat{a})\chi_\pm (\vec{a}) = \pm \chi_\pm (\vec{a})\ .
\label{eq11}
\end{equation}

It is now a simple matter to extend these projection operators to projections in 
$2^N$-dimensional vector spaces.   In general we may write

\begin{equation}
\hat{P}_{s_1,s_2,...s_N}(\hat{a}_1,\hat{a}_2,... \hat{a}_N) = 
\bigotimes_{i = 1}^N  \hat{P}_{s_i}(\hat{a_i})\ ,
\label{e12}
\end{equation}

for $s_i \in \{\pm\}$. It is easy to see that these projection operators satisfy 
our orthogonal idempotent constraints.   We then define our set of eigenvectors 
in a similar manner

\begin{equation}
\chi_{s_1,s_2,...s_N}(\vec{a}_1,\vec{a}_2,... \vec{a}_N) = \bigotimes_{i = 1}^N  
\chi_{s_i}(\vec{a_i})\ .
\label{eq13}
\end{equation}

By restricting our considerations to the vector space of spinors we are able to give these 
abstract considerations physical significance.   Recall that the rotation operator corresponding 
to a rotation of $\theta$ about the axis defined by $\hat{a}$ for the vector space of 
two-component spinors takes the form 

\begin{equation}
\begin{aligned}
\hat{R}(\theta,\hat{a})&=\cos{\frac{\theta}{2}} + i (\vec{\sigma}\cdot\hat{a})\sin{\frac{\theta}{2}} \\
&= e^{\frac{i\theta}{2}}\hat{P}_{+}(\hat{a}) + e^{-\frac{i\theta}{2}}\hat{P}_{-}(\hat{a})\ .
\end{aligned}
\label{eq14}
\end{equation}

It follows that eigenstates of our projection operators $\hat{P}_{\pm}(\hat{a})$ are physically 
invariant under rotations about $\hat{a}$, differing only by a phase.  We therefore identify 
$\hat{P}_{\pm}(\hat{a})$ as projecting out the portion of the state vector with spin parallel 
($+$) or anti-parallel ($-$) to the $\hat{a}$-axis.  By choosing $\vec{a}=\vec{p}$, 
where $\vec{p}$ is the three-momentum of the particle, we find $\hat{P}_\pm(\hat{p})$ 
to be the helicity projection operators.  This most neatly encapsulates the experimental 
observance of binary spin states in mathematical terms.

We now wish to use our projection operator methodology to classify states of positive 
and negative energies, e.g. matter/anti-matter. The inclusion of an additional two-valued 
quantum property necessitates (at minimum) a four-dimensional vector space.   We 
therefore construct the projection operators of the form

\begin{equation}
\hat{P}_{s_1,s_2}(\hat{q},\hat{p}) = \hat{P}_{s_1}(\hat{q}) \otimes  \hat{P}_{s_2}(\hat{p})\ ,
\label{eq15}
\end{equation}
where we have introduced the vector $\hat{q}$ about which we will have more to say shortly, 
for the time being $\hat{q}$ is simply a set of three complex numbers and satisfies 
$\hat{q}\cdot\hat{q}=1$.  

Next we define the operand

\begin{equation}
\chi_{s_1,s_2}(\hat{q},\hat{p}) = \chi_{s_1}(\hat{q}) \otimes  \chi_{s_2}(\hat{p})\ .
\label{eq16}
\end{equation}
The corresponding generalization of Eq. (\ref{eq11}) yields

\begin{align}
(\vec{\sigma}\cdot\hat{q}\otimes I_2) \chi_{s_1,s_2}(\hat{q},\hat{p}) = s_2 
\chi_{s_2,r_2}(\hat{q},\hat{p})\\
(I_2\otimes \vec{\sigma}\cdot\hat{p}) \chi_{s_1,s_2}(\hat{q},\hat{p}) = s_1 
\chi_{s_1,s_2}(\hat{q},\hat{p})\ .
\label{eq17}
\end{align}
Exploiting the fact that $s_1=\pm s_2$ we may construct the equations

\begin{align}
(\vec{\sigma}\cdot\hat{q}\otimes I_2 + I_2\otimes \vec{\sigma}\cdot\hat{p}) 
\chi_{\pm,\mp}(\hat{q},\hat{p})=(\vec{\sigma}\cdot\hat{q} \oplus \vec{\sigma}
\cdot\hat{p})\chi_{\pm,\mp}(\hat{q},\hat{p}) = 0\\
(\vec{\sigma}\cdot\hat{q}\otimes I_2 - I_2\otimes \vec{\sigma}\cdot\hat{p})
\chi_{\pm,\pm}(\hat{q},\hat{p})=(\vec{\sigma}\cdot\hat{q}\ominus \vec{\sigma}
\cdot\hat{p}) \chi_{\pm,\pm}(\hat{q},\hat{p})=0\ .
\label{e18}
\end{align}

We now identify the set of gamma matrices in the chiral representation

\begin{equation*}
\begin{aligned}
\gamma^{0} &= \sigma^{1} \otimes I_2 \\
\gamma^{k} &= i\sigma^{2} \otimes \sigma^{k} 
\end{aligned}
\label{eq19}
\end{equation*}
satisfying the Clifford algebra

\begin{equation*}
\begin{aligned}
\{\gamma^{\mu},\gamma^{\nu}\} &= 2\eta^{\mu \nu} I_4\ .
\end{aligned}
\label{eq20}
\end{equation*}

Taking the usual definition $\gamma^{5}=i\gamma^{0}\gamma^{1}\gamma^{2}\gamma^{3}$ 
as the matrix which anticommutes with all $\gamma^\mu$, let us write

\begin{equation*}
\begin{aligned}
i\gamma^{5} \gamma^{0} &= \sigma^{2} \otimes I_2 \\
-\gamma^{5} &= \sigma^{3} \otimes I_2 
\end{aligned}
\label{eq21}
\end{equation*}

which give

\begin{align}
(\vec{\sigma}\cdot\hat{q}\otimes I_2) &= \frac{1}{|\vec{q}|} \gamma^0 \gamma^5 
(\gamma^5 q_1 - i I_4 q_2 +\gamma^0 q_3) \\
(I_2 \otimes\vec{\sigma}\cdot\hat{p}) &=- \frac{1}{|\vec{p}|} \gamma^0 \gamma^5 
\vec{\gamma}\cdot \vec{p}\ .
\label{eq21}
\end{align}
Rewriting the operators of Eqs. (\ref{eq15}) and (\ref{eq16}) in terms of the gamma matrices, we find

\begin{align*}
(\vec{\sigma}\cdot\hat{q}\oplus \vec{\sigma}\cdot\hat{p})=\frac{1}{|\vec{p}||\vec{q}|} 
\gamma^0 \gamma^5 (\gamma^0 |\vec{p}|q_3- |\vec{q}|\vec{\gamma}\cdot \vec{p}+
\gamma^5 |\vec{p}|q_1 - i I_4 |\vec{p}|q_2)\\
(\vec{\sigma}\cdot\hat{q}\ominus \vec{\sigma}\cdot\hat{p}) =\frac{1}{|\vec{p}||\vec{q}|} 
\gamma^0 \gamma^5 (\gamma^0 |\vec{p}|q_3+ |\vec{q}|\vec{\gamma}\cdot \vec{p}+
\gamma^5 |\vec{p}|q_1 - i I_4 |\vec{p}|q_2 )\ .
\end{align*}

It is our goal to identify the vector $\vec{q}$ with our physical quantities.  Scaling these operators from the left with $\pm|\vec{p}||\vec{q}| \gamma^5\gamma^0$, we now 
restrict $\vec{q}$ and $\vec{p}$ to be of equal-magnitude (this is equivalent to the on-mass-shell asssumption). We then obtain equivalent equations to Eqs. (\ref{eq15}) 
and (\ref{eq16}) of the form

\begin{align}
(-\gamma^0 q_3 + \vec{\gamma}\cdot \vec{p}+\gamma^5 q_1 - i I_4 q_2)\chi_{\pm,\mp}&=0 \label{eq23} \\
(+\gamma^0 q_3 + \vec{\gamma}\cdot \vec{p}-\gamma^5 q_1 + i I_4 q_2)\chi_{\pm,\pm}&=0\ . \label{eq24}
\end{align}

It is now possible to make the identifications with our physical quantities explicit.  For the 
vector $\vec{q}$, we identify:

\begin{equation}
q_1 = i m \sin{\alpha} \qquad
q_2 = -i m \cos{\alpha} \qquad
q_3 = p_0 = E
\label{eq25}
\end{equation}

The identification of $q_3=E$ ($E>0$) is chosen to align with our definition of gamma matrices though equivalent linear combinations of the vector may be found through unitary transformations. It may appear curious upon first inspection that the vector $\vec{q}$ is seemingly compelled to take on imaginary values in two components and real values in the third. It is, however, a simple matter to absolve ourselves of this inhomogeneity by performing a Wick-rotation of the energy axis in the complex plane and thereby considering the four vectors of Minkowski spacetime in purely Euclidean terms. In this way the connection between our projection operators as a basis of $SL(2,\mathbb{C})$ and the Lorentz group $SO(1,3)$ may be made explicit. These simplifications notwithstanding, we will continue to consider real-valued energies in Minkowski spacetime. 

Substituting our terms of Eqs. (\ref{eq25}), into Eqs. (\ref{eq23}) and (\ref{eq24}) we  obtain

\begin{align}
\label{Eq26}
(-\gamma^0 E + \gamma^k p_k + m e^{i \alpha \gamma^5})\chi_{\pm,\mp}&=0 \\
\label{Eq27}
(+\gamma^0 E +\gamma^k p_k - m e^{i \alpha \gamma^5})\chi_{\pm,\pm}&=0\ .
\end{align}
It is clear that Eqs. (\ref{Eq26}) and (\ref{Eq27}) are the momentum space analogues of the CDE for positive and negative energies and therefore are equivalent to plane wave solutions of Eq. (\ref{eq3}). The eigenstates therefore satisfy

\begin{align}
E^2 \chi_{s_1,s_2}=(p^2+m^2)\chi_{s_1,s_2}\ .
\end{align}

The method of projection operators for deriving fundamental equations has potential to extend beyond the $\mathbb{C}^4$ vector space of bi-spinors. While outside the scope of the present work, the possibility of extending the concepts and methodologies presented here to investigate the algebraic structure inherent in three particle flavors and their mass spectrum remains a tantalizing possibility.

\section{Symmetries and their physical implications}

\subsection{Spacetime symmetries}

From the Special Theory of Relativity we know that in order for an equation to be considered fundamental it must remain invariant in all inertial frames of reference. Such frames may be defined as those frames in which the symmetries of Minkowski spacetime are agreed to hold. These symmetries are represented by the Poincar\'{e} group $\mathcal {P}\ =\ SO (3,1) \otimes_s T (3+1)$ consisting of rotations, boosts, and translations in space and time (see Section 2.1) which carry one inertial frame to another. We may therefore refer to the class of inertial observers to whom fundamental equations must remain invariant as as Poincar\'{e} observers. It follows as a neccesary condition for any equation to be considered fundamental that it preserve its form for all Poincar\'{e} observers and hence remain invariant with respect to all transformations given by $\mathcal {P}$. Only such equations will be able to make physical predictions about which all Poincar\'{e} observers will agree upon. It is easy to verify that the derived CDE above is one such Poincar\'e invariant equation and so satisfies a necessary condition to be called a {\it fundamental equation of physics}. 

The other characteristics requisite of a fundamental theory may be summarized as locality, gauge invariance, and the satisfaction of a least-action principle (equivalent to the existence of Lagrangian density for the equation).  Since the CDE is a first-order partial differential equation, it is local.  In this paper only free particles are considered,
therefore, gauge invariance is not included.  In Section 2.2, we demonstrated that the Lagrangian density for the CDE exists and may be written in the fully symmetric form given by Eq. (\ref{eq6}).  In general Lagrangians posses less 
symmetry than the dynamical equations resulting from them due to the assumptions on which the Noether theorem is based [18,19].  The best known example is the law of inertia, whose dynamical equation is Galilean invariant but its Lagrangian is not [20,21].  However, a method to restore Galilean invariance of the Lagrangian was developed and applied to the law of inertia [22].

We followed [17] to construct the Dirac Lagrangian, which is already Poincar\'e invariant.  Similarly, the Lagrangian for the CDE equation is also Poincar\'e invariant. Therefore the form of the CDE equation and its Lagrangian, as well as theoretical predictions resulting from them, are the same for all observers who accept the Principle of Relativity underlying Special Theory of Relativity and its Poincar\'e group $\mathcal {P}$. Therefore, the presented results combined with the above discussion clearly imply that any theory of physics based on the CDE equation and its Lagrangian may be rightly called a fundamental equation of physics. 

\subsection{CPT symmetries}

Beyond the symmetry group of $\mathcal{P}$, we may also inquire into which discrete symmetries of nature hold for the CDE. We define the operations of charge-conjugation 
($\mathcal{C}$), spatial inversion ($\mathcal{P}$), and time-reversal ($\mathcal{T}$) 
on an arbitrary operator $\hat{O}(x,t)$ acting on a state $\psi(x,t)$ as follows

\begin{align}
\mathcal{C}[\hat{O}(x,t)\psi(x,t)]\mathcal{C}^{-1} &= \hat{O}^*(x,t)\psi^*(x,t) \\
\mathcal{P}[\hat{O}(x,t)\psi(x,t)]\mathcal{P}^{-1}&=\hat{O}(-x,t)\psi(-x,t) \\ 
\mathcal{T}[\hat{O}(x,t)\psi(x,t)]\mathcal{T}^{-1} &= \hat{O}^*(x,-t)\psi(x,-t)\ .
\label{eq27}
\end{align}

With these definitions, we may determine the parity-conjugated forms of the CDE. Defining $\hat{D}(\alpha)= i \gamma^{\mu} 
\partial_\mu - m e^{i\alpha \gamma} $, taking the symmetry operation $\mathcal{U} \in \{ \mathcal{C},\mathcal{P},\mathcal{T} \}$, and writing the corresponding unitary transformation acting on the spinor indices $\hat{U} \in \{ \hat{C},\hat{P},\hat{T} \}$, 
we obtain

\begin{align}
\hat{U}[\mathcal{U}\hat{D}(\alpha)\mathcal{U}^{-1}] \hat{U}^\dagger \psi_U(x,t) &= 0 
\end{align}
where $ \psi_U(x,t) \equiv \hat{U} [\mathcal{U} \psi(x,t) \mathcal{U}^{-1}]$. Then, the resulting transformations may be summarized as

\begin{align}
\hat{C}[\mathcal{C}\hat{D}(\alpha)\mathcal{C}^{-1}] \hat{C}^\dagger &= \hat{D}(\alpha^*) \\
\hat{P}[\mathcal{P}\hat{D}(\alpha)\mathcal{P}^{-1}] \hat{P}^\dagger &= \hat{D}(-\alpha) \\
\hat{T}[\mathcal{T}\hat{D}(\alpha)\mathcal{T}^{-1}] \hat{T}^\dagger &= \hat{D}(-\alpha^*)\ ,
\label{eq28}
\end{align}
and the following conditions can be identified\\
1. If $\psi$ exhibits C-invariance ($\psi=\psi_C$), then $\alpha \in \mathbb{R}$  

2. If $\psi$ exhibits CP-invariance ($\psi=\psi_{CP}$), then $\alpha \in \mathbb{I}$ 

3. If $\psi$ exhibits CPT-invariance ($\psi=\psi_{CPT}$), then $\alpha \in \mathbb{C}$ \\

These conditions combined with the CPT theorem reinforce our conclusion that the Dirac  equation with chiral freedom is the most general first order differential equation derivable from the irreps of the Poincar\'{e} group. While it is quite conceivable that the constraints imposed by observed classes of discrete symmetries in interactions (and their fundamental violations) may contribute to constraints on the CDE, it is important to emphasize the regimes of validity presented here are derived for free particles in the absence of interactions and therefore a result of extrinsic and not intrinsic symmetries. In this way the discrete constraints presented above are foundational for any extended theoretical considerations.   

\section{Conclusions}

We have presented three distinct derivations of the generalized form of Dirac's equation to include all those parameters not constrained by Poincar\'{e} symmetry: from the irreducible representations of the Poinccar\'{e} group, from Lagrangian formalism, and from a classification scheme built from projection operators. The result of all these derivations agree and show the chiral angle is a necessary degree of freedom for complete categorization of all physical eigenstates of spin and the matter/anti-matter nature. Equivalently, by showing that the obtained chiral Dirac equation is local, Poincar\'e invariant and its Lagrangian exists, the equation becomes a fundamental equation of modern physics.  Using this derived equation it has then possible to investigate the underlying spacetime and CPT symmetries and the constraints on allowable values. 

\bigskip\noindent
{\bf Author Contributions:}
Conceptualization, T.B.W. and Z.E.M..; 
methodology, T.B.W. and Z.E.M.; validation, T.B.W. and Z.E.M.; 
formal analysis, T.B.W.; investigation, T.B.W. and Z.E.M.; 
writing---original draft preparation, T.B.W. and Z.E.M.; 
writing---review and editing, T.B.W. and Z.E.M.  Both authors 
have read and agreed to the published version of the manuscript.

\bigskip\noindent
{\bf Funding:}
This research received no external funding.

\bigskip\noindent
{\bf Instututional Review Board Statement:}
Not applicable.

\bigskip\noindent
{\bf Informal Consent Statement:}
Not applicable.

\bigskip\noindent
{\bf Data Availability Statement:}
Not applicable. 

\bigskip\noindent
{\bf Acknowledgments}
We are grateful to two reviewers for reading our paper and providing 
comments and suggestions for its improvement.  Our special thanks to 
M. Efroimsky (one of the reviewers) for bringing to our attention 
differences between our derivation and that presented by Ryder [15]. 

\bigskip\noindent
{\bf Conflict of Interest:}
The authors declare no conflict of interest.


\begin{thebibliography}{999}

\bibitem{1} P.A.M. Dirac, The quantum theory of the electron, Proc. 
                  Royal Soc. London 117 (1928) 610
\bibitem{2} I. Sogami, Prog. Theor. Phys. 66 (1981) 303
\bibitem{3} S.I. Kruglov, arXiv:hep-ph / 0507027v2 23 June 2006
\bibitem{4} E. Marsh, Y. Narita, Front. Phys. 3 (2015) Article 82
\bibitem{5} A.O. Barut, P. Cordero, G.C. Ghirardi, Phys. Rev. 182 (1969) 1844
\bibitem{6} A.O. Barut, The mass of muon, Phys. Lett. 73B (1978) 310
\bibitem{7} W. Pfister, Mixed-symmetry solutions of generalized three-particle Bargmann-Wigner
                  equations in the strong-coupling limit, Nuovo Cimento A 108 (1995) 1427
\bibitem{8} S.I. Kruglov, On the Hamiltonian form of generalized Dirac equation for 
                  fermions with two mass states, Elect. J. Theor. Phys. 3 (2006) 11
\bibitem{9} S.I. Kruglov, Modified Dirac equation with Lorentz invariance violoation
                  and its solutions for particles in an external magnetic field, 
                  Phys. Let. B  718 (2012) 228
\bibitem{10} D. Leiter, G. Szamosi, Pseudoscalar mass and its relationship to 
                   conventional scalar mass in relativistic Dirac theory of the electron, 
                   Let. Nuovo Cim. 5 (1972) 814
\bibitem{11} K. Nozari, Generalized Dirac equation and its symmetries, 
                   Chaos, Solitons \& Fractals 32 (2007) 302
\bibitem{12} T.B. Watson, Z.E. Musielak, Chiral symmetry in Dirac equation and 
                    its effecst on neutrino masses and dark matter, Int. J. Mod. Phys. A 
                    35 (2020) 2050189
\bibitem{13} E.P. Wigner, On unitary representations of the inhomogeneous Lorentz group, 
                   Ann. Math. 40 (1939) 149
\bibitem{14} Y.S. Kim and M.E. Noz, Theory and Applications of the 
                  Poincar\'e Group, Reidel, Dordrecht, 1986
\bibitem{15} L.W. Ryder, Quantum Field Theory, Cambridge University Press,
                  Cambridge, 1985
\bibitem{16} P.H. Frampton, Gauge Field Theories, John Wiley \& Sons, Inc., 
                  New York, 2000
\bibitem{17} N.A. Daughty, Lagrangian Interactions, Addison-Wesley Publ. Comp., Inc.,
                   Sydney, 1990
\bibitem{18} S. Hojman, Symmetries of Lagrangians and of their equations of motion, 
                   J. Phys. A: Math. Gen. 17 (1984) 2399
\bibitem{19} S. Hojman, A new conservation law constructed without using either 
                   Lagrangians or Hamiltonians, J. Phys. A: Math. Gen. 27 (1992) L59
\bibitem{20} L.D. Landau and E.M. Lifschitz, Mechanics, Pergamon Press, Oxford, 1969
\bibitem{21} J.-M. Levy-Leblond, Group-theoretical foundations of classical mechanics:
                    the Lagrange gauge problem, Comm. Math. Phys. 12 (1969) 64
\bibitem{22} Z. E. Musielak and T. B. Watson, gauge functions and Galilean invariance 
                   of Lagrangians, Phys. Let. A 384 (2020) 126642
\bibitem{23} M. Petitjean, About chirality in Minkowski spacetime, Symmetry, 11 (2019) 1320

\end{thebibliography}
\end{document}